# The doping and disorder dependent variation of the isotope exponent in hole doped cuprates: A non-superconducting perspective

S. H. Naqib* and R. S. Islam

*Department of Physics, University of Rajshahi, Rajshahi-6205, Bangladesh*

Since the early days of the discovery of hole doped high-$T_c$ cuprates, the variation of the oxygen isotope exponent (IE) with the number of doped holes, $p$, in the $CuO_2$ planes has been a source of considerable debate. There is a growing acceptance in the high-$T_c$ community that the mediating bosons leading to Cooper pairing have largely electronic character. At the same time there is an increasing acknowledgement that at some level the lattice-charge interactions might be important in cuprates. The experimentally observed substantially large IE over certain doping range always casts a shadow over any proposal where non-phononic mechanism is invoked. Besides, the various existing theoretical schemes, based on electron-phonon interactions, cannot describe the anomalous features shown by the IE as a function of hole concentration/disorder, either quantitatively or qualitatively. Based on a recent experiment relating the possibility of Fermi-surface reconstruction to the thermoelectric transport measurements (Nature Commun. **2:432** doi: 10.1038/ncomms1440 (2011)), we propose here a simple scenario where isotope substitutions affect the charge/spin *stripe state* via the coupling to the underlying lattice and thereby change the superconducting $T_c$. In this picture, significant part of the IE, over an extended $p$-range, actually originates from the isotope induced stripe modulation and is not directly related to the characteristic energy scale of the pairing phonons as is the case in conventional BCS theory. Our proposal qualitatively explains all the disorder- and $p$-dependent features of the IE seen in hole doped high-$T_c$ cuprates. We also provide with an outline of some experiments that can verify the degree of validity of the proposed scenario.

Ever since the discovery of the hole doped high-$T_c$ cuprates [1], the question of how superconductivity emerges from the parent antiferromagnetic (AFM) insulator upon adding extra holes ($p$) in the $CuO_2$ plane has been of immense interest. After more than a quarter of a century, the answer to this question still remains elusive [2 – 4]. It is widely accepted that the normal and superconducting (SC) state properties of these strongly correlated electronic materials are non-Fermi liquid like [5]. The *T-p* phase diagram shows an intricate interplay among a number of electronic ground states (*e.g.*, AFM Neel order, spin glass state, stripe correlations, pseudogap (PG) region, SC phase, and the Fermi-liquid like metallic state as one increases the hole content) and it is extremely hard to formulate a coherent theoretical scheme describing all these different facets [5 – 7] as the doping levels are varied. Not surprisingly, the conventional BCS formalism [8], developed for weakly interacting metallic systems, does not describe various features shown at and below the SC transition temperature in cuprates.

The isotope effect, which played a pivotal role in the development of the BCS formalism within the framework of electron-phonon (e-ph) interaction, is highly complicated in the case of cuprate superconductors [9 – 11]. Whether phonons play a significant role as the mediating bosons for Cooper pairing in hole doped cuprates is a matter of strong controversy [9 – 13]. The magnitude of the IE in cuprates depends strongly on the number of doped holes in the $CuO_2$ planes and therefore, may indicate that phonons play different quantitative roles as the mediating boson at different regions of the $T_c$-*p* phase diagram. The general trend being, $α(p)$ is large inside a certain doping range in the underdoped (UD) side, and can greatly exceed the canonical BCS value of 0.5 near the so-called $1/8^{th}$ anomaly [9, 10, 14, 15], the value of the IE near the optimum doping (OPD), on the other hand, is quite small and it stays almost the same in the overdoped (OD) region [9, 10, 15]. The question that naturally arises is that whether a larger/smaller value of the IE corresponds to a greater/lesser contribution of the phonons in the pairing mechanism. No definitive answer exists so far.

Unlike in the BCS superconductors, the interpretation of the strongly *p*-dependent IE in cuprates poses a serious theoretical challenge because of the presence of various electronic correlations (strengths of which also depend on the hole content). As long as the interplay of these correlations among themselves, and more importantly with superconductivity itself is unclear, the significance of $α(p)$ will be hard to comprehend. For example, the IE is affected by the pseudogap (PG), spin/charge ordering (stripe correlations), and possibly by other exotic density waves in the quasi-particle (QP) energy spectrum [15 – 17]. There exists a number of theoretical schemes aimed to describe the doping and disorder dependence of the IE in hole doped cuprates [9, 15, 18, 19] but none of these schemes explain the doping and disorder dependent IE over the entire range of hole contents [10].

An unprecedentedly high-$T_c$, proximity to the AFM order, and the *d*-wave order parameter have led the theorists to put forward a number of non-phononic mechanisms for hole doped cuprates [20 – 23]. Especially, there has been a growing prominence of pairing theories based on spin fluctuations [21 – 23]. Recently optical spectroscopic study [24] on Bi2212 crystals have revealed the importance of the role of non-phononic electronic excitations (AFM spin fluctuations) in Cooper pairing. At the same time, results from resonant inelastic X-ray scattering [21] uncovered the presence of intense paramagnon excitations in Y123 and Y124 compounds over a wide range of hole contents. How one reconciles the apparently contradictory significant IE with the findings from these experimental probes [21, 24] poses a serious conceptual problem. In this short communication we wish to address this issue by invoking solely to the stripe correlations.

Besides superconductivity and the PG correlations, the spin/charge ordered *stripe state* [7] is probably the most widely studied topic in the field of research on high-$T_c$ cuprates. Experimentally, the static spin/charge stripe correlations are found in UD cuprates in the vicinity of $p \sim 0.125$ (the so-called $1/8^{th}$ anomaly) [7, 16], although dynamical (fluctuating) stripe correlations are believed to exist over a much wider doping range, especially in the single $CuO_2$ layer LSCO compounds [7]. Incommensurate low-energy spin fluctuations, generally interpreted as precursor to strong stripe correlations, are also observed in double-layer Y123 and Bi2212 [7, 25, 26]. These days stripe correlations are widely considered as a generic feature of all the hole doped cuprate superconductors over a significant region of the *T-p* phase diagram. The possible link between stripe and SC correlations are not completely understood at this moment but it is fair to say that static/stabilized stripe order competes with superconductivity [7, 27] and reduces the superconducting transition temperature. Stripes can be described as unidirectional density wave states, which modulate both charge and spin densities in the real space. The precise mechanism of stripe formation in a particular system is not entirely clear yet. Charge/spin stripe probably forms in doped Mott insulators as a compromise between the AFM ordering among the Cu spins and strong Coulomb repulsion between the electrons (both favoring localization) with the kinetic energy of the mobile doped carriers (leading to delocalization). Static stripe order breaks the underlying crystal lattice symmetry and the resulting charge order is particularly strongly coupled with the lattice degrees of freedom. For dynamical stripe correlations, the coupling is weaker but still sufficiently strong to affect each other. In the stripe phase there is ionic displacements due to charge ordering. In such a situation any change in the lattice affects the charge order and vice versa. In fact the detection of stripe correlations by neutron diffraction becomes possible because of this strong coupling between lattice and charge degrees of freedom. Therefore, it is not surprising that location and motion of particular atomic species (*e.g.*, the apical oxygen in La214) can have significant effect on stabilization of stripe order [28].

The significance of spin/charge ordered states in the physics of hole doped cuprates have been demonstrated by recent comparative study of the thermoelectric transport in Y123 and Eu-LSCO systems [29]. This work [29] also suggests boldly that the Fermi-surface reconstruction, as seen by quantum oscillation experiments [30] at low temperatures, is possibly due to the breaking of the lattice translational symmetry caused by the stripe ordering. The appearance of field induced charge ordering in Y123 [31] lends further support to the assumption that stripe order is an essential ingredient for the understanding of these strongly correlated electronic systems. The thermoelectric transport properties also showed that the basic features of the stripe related phase diagrams for Y123 and Eu-LSCO are essentially the same and the charged stripe phase exists over an extended doping range from $p = 0.08$ to $0.18$ [29] in both the systems. By taking into consideration of these [29, 31] experimental results, we put forward a simple scenario as an explanation for the $p$-dependent anomalous IE observed in cuprates as follows. Isotopic substitution changes the lattice dynamics even if only by changing the characteristic phonon energies. In a situation where lattice and QP degrees of freedoms are strongly intertwined, any change in the phonon spectrum will modify the stripe order and consequently change $T_c$, as the degree of stripe stabilization determines the degree of competition between the two correlations. In fact direct experimental evidence in favor of such a proposal can be found in isotope effect experiments on Nd-LSCO [14]. In general, the stronger the stripe ordering, the stronger would be the coupling between the charge and lattice degrees of freedom and consequently the effect of isotope substitutions would be more pronounced. It is important to note that in this picture isotopic substitution slows down the lattice vibration (*e.g.* the case of $O^{16} \rightarrow O^{18}$ substitution) and thereby reduces fluctuations of the strongly coupled stripe order. Within this scenario the observed reduction in $T_c$ results from a relatively more stabilized stripe order in the isotope substituted compound and is not directly linked to the change in the energy scale of the characteristic harmonic phonons as is envisaged in the conventional BCS formalism. This proposal explains quite naturally the anomalous increase of the IE in the vicinity of the $1/8^{th}$ doping. At this hole doping stripes are maximally stabilized irrespective of the different families of hole doped cuprates and therefore the charged QP-lattice coupling is at its strongest. Consequently any change in the phonon spectrum affects the striped state most significantly around the $1/8^{th}$ anomaly, thereby inducing the largest shift in $T_c$ upon isotopic substitution. To illustrate the above points, we show the *p*-dependent oxygen ($O^{16} \rightarrow O^{18}$) IE for LSCO (details can be found in refs. [10, 15, 32]) together with the characteristic temperature, $T_{s0}$, taken from ref. [29], in Fig. 1. $T_{s0}$ is the temperature where the normal state Seebeck coefficient changes its sign due to a reconstruction of the Fermi-surface as the stripe correlations set in and break the translational symmetry [29]. The correspondence between the doping evolutions of $T_{s0}(p)$ and $\alpha(p)$ is striking. To our knowledge, Fig. 1 shows the first clear systematic link between $\alpha(p)$ with any other *p*-dependent parameter relevant to the physics of hole doped cuprates.

It is important to note that $\alpha(p)$ remains small but finite outside the *p* range from 0.08 to 0.18 where stripe correlations either do not exist or are severely weakened. The small IE in these doping regions could be due to a *truly phonon mediated pairing mechanism*. In fact a relatively small e-ph coupling constant ~ 0.4, was found by Conte *et al.* [24], whereas the coupling constant due to spin fluctuations (e-sf) were found to be ~ 1.1. In the case of a joint pairing mechanism, where the phonons only play a minor role, a greatly reduced IE is expected. In its simplest form, the IE in the case of mixed mechanism [9, 33], can be expressed as [9, 34], $\alpha = 0.5 [1 - \lambda_I/(\lambda_I + \lambda_{e-ph})]$, where $\lambda_I$ is the non-phononic (e.g., $\lambda_{e-sf}$) coupling constant and $\lambda_{e-ph}$ is the coupling constant due to the e-ph interaction. Using the values obtained by Conte *et al.* [24], the pairing related IE from phonons turns out to be ~ 0.13, a value not far away from the experimentally observed $\alpha(p)$ outside the *p*-range from 0.08 to 0.18. Irrespective of the different variants of the pairing models, addition of the effect of Coulomb pseudopotential always reduces the value of the IE even further. Cuprates are systems with low carrier density and the effect of poorly screened Coulomb repulsion is bound to play a role. This in turn implies that the pairing related IE in cuprates is masked by the strong stripe related effects over a significant doping range.

Quite interestingly, the proposed scenario also explains the disorder induced enhancement in IE [9, 10] qualitatively. For example, it has been observed that Zn pins stripe fluctuations [7, 35] and at the same time at a given *p*, IE increases in Zn substituted samples [9, 10, 15]. In view of the present proposal, IE gets enhanced because the isotopic substitution had a greater effect in modifying the relatively more stabilized stripe ordered state in the Zn doped compounds compared to the Zn-free ones. In the OD side Zn has a much smaller effect on the IE [10] because the stripe orders are already severely weakened or even absent and Zn is unable to have a significant effect on the stripe dynamics. As an exception to the rule, Zn substitution is seen to diminish the maximally enhanced $\alpha(p)$ at found in Zn-free samples at $p \sim 0.125$ [15]. At this precise doping the stripe order in LSCO is at its strongest and Zn plays no further appreciable role in pinning. Instead, the spin vacancies created by nonmagnetic Zn degrades the integrity of the stabilized stripe order [36, 37]. Ni substitution, on the other hand, has almost no effect on the IE in LSCO [38]. This is puzzling since both Zn and Ni act as pair-breakers [38] and should enhance the IE in same qualitative manner [9]. In relation to this observation, we would like to draw attention of the neutron scattering experiments on Zn and Ni substituted LSCO by Kofu *et al.,* [39], where it was found that Zn and Ni affect the low energy spin excitations in LSCO quite differently. While there was clear indication that Zn slowed down spin fluctuations, Ni failed to do so [39]. This indicates that nonmagnetic Zn and magnetic Ni play vastly different roles in stabilizing the stripe correlations in cuprates and offers a ready explanation for the ineffectiveness of Ni to enhance the IE in LSCO. This might also have a possible relevance to the interesting fact that unlike Ni, Zn suppresses $T_c$ much more effectively.

A question might be raised regarding the degree of validity of comparing $T_{s0}(p)$ of Y123 and Eu-LSCO with the $\alpha(p)$ for pure LSCO. To address this issue, we would like to mention the following facts. Pure LSCO is a system which has been confidently characterized as stripe ordered or nearly stripe ordered (*liquid*) over a wide range of Sr contents [7]. Nd/Eu is doped in LSCO only to amplify the effects of stripe correlations by stabilizing the order via anisotropic lattice distortion [40]. Our proposal predicts a further systematic enhancement in the $\alpha(p)$ in Nd/Eu-LSCO compared to that found in pure LSCO, as confirmed partly in ref. [14]. It is worth mentioning that the dynamic magnetic correlations in Nd/Eu doped and pure LSCO are essentially identical [40]. $T_{s0}(p)$ data for Y123 is included in Fig. 1 to illustrate the possible universality of the proposed scenario. Y123 with its double $CuO_2$ planes, $CuO_{1-\delta}$ chains, and much larger $T_c$, is a very different system from LSCO. Nevertheless, the $T_{s0}(p)$ data for LSCO and Y123 are almost indistinguishable.

At this point we would like to put forward a list of possible experiments that can check the validity of some specific predictions that follows directly from the proposed picture. (i) The stripe order *melts* at higher temperatures due to thermal fluctuations. Within this context of stripe melting, we expect a noticeable effect on $T_{s0}/T_{c0}$ [29] of isotopic substitution ($T_{c0}(p)$ being the characteristic charge ordering temperature). $T_{s0}(p)/T_{c0}(p)$ should increase systematically as the lower mass atomic species is replaced by the higher ones. (ii) A particularly interesting system should be the Zn doped Eu/Nd-LSCO with $p \sim 0.125$. At this composition the Zn-free compound is maximally stripe ordered. Zn substitution will primarily create spin vacancies and will have negligible effect on stripe stabilization. We predict a significantly diminished IE in such a system. As mentioned earlier, somewhat similar effect was observed in Zn doped LSCO with $p \sim 0.125$ [15]. (iii) In Nd-LSCO and related systems charge and spin ordering is intimately linked with LTT to LTO structural transition [7, 40]. A systematic study of the possible variation of structural transition temperature with isotope substitution can clearly elucidate the isotope induced change in the stripe dynamics. (iv) So far, detailed isotope effect experiments were mainly done on systems related to Y123 and LSCO. We expect that the $\alpha(p)$ trend should be very similar to those shown in Fig. 1 in other hole doped systems, even though a quantitative difference should exist as the degree of stripe correlations vary from system to system.

It should be mentioned that the degree of incommensuration ($\delta$) found in the inelastic neutron scattering experiments depends on the hole content in the $CuO_2$ plane. Quite interestingly, Yamada *et al.*, [41] found an empirical relation between $\delta(p)$ and $T_c(p)$, given by $k_B T_c(p) = \hbar v^* \delta(p)$. Based on this relation Rosengren *et al.* [18] have analyzed the isotopic shift assuming a Josephson coupled stripe order where $T_c$ is changed via the change in $v^*$. The parameter $v^*$ might have some relevance with stripe dynamics (it has the dimension of velocity) but its physical meaning in unclear as its magnitude differs widely from the Fermi or spin density wave velocities. Rosengren *et al.*, found that a change in $v^*$ by 5% can account for the experimentally observed oxygen IE at certain hole contents in Y123 and LSCO. On the other hand, greatly enhanced IE in Nd-LSCO compounds led Wang *et al.*, [14], to conclude that e-ph interaction is essential for Cooper pairing. In this paper, we are offering a completely different point of view. It is noteworthy that the charge/spin ordering temperature goes to zero at $p = 0.08$ and $0.18$ in the UD and OD sides, respectively. These two *p*-values are tantalizingly close to the two special values of hole contents where possible quantum critical points (QCPs) are located. The UD one is related to a metal-insulator transition [42], while the one in the OD side, has non-Fermi liquid metallic character, the PG vanishes quite abruptly at this doping [43 – 45], and the Fermi surface undergoes a fundamental change due to a *p*-induced change in the chemical potential [46].

In the absence of a recognized theory for the pairing in the hole doped cuprates, it is important to search for universal trends and correlations among different physical parameters as guides to the essential physics for these strongly correlated electronic systems. Fig. 1 provides us with such a correlation. A clear correspondence between $T_{s0}(p)$ and $\alpha(p)$ supplies the strongest circumstantial evidence in favor of the proposed scenario.

Before concluding, we would like to stress that this paper does not offer any theory neither even a framework. The possible value of this work lies in an observation linking two apparently disjointed experimental observations that might have some implications in the development of a pairing theory for hole doped cuprate superconductors.

To summarize, we have put forward a scenario where the isotope effect in hole doped cuprates over an extended range of hole contents is dominated by the interplay between stripe and SC correlations. Our proposal provides with a way to reconcile the significant IE with non-phononic pairing in high-$T_c$ cuprates. This simple picture qualititatively explains all the doping and disorder dependent features of IE in hole doped cuprates.

The authors acknowledge the Commonwealth Commission, UK, Trinity College, University of Cambridge, UK, the Industrial Research Limited and MacDiarmid Institute for Advanced Materials and Nanotechnology, Wellington, New Zealand, for financial support and for providing with the experimental facilities. The authors also acknowledge the AS-ICTP, Trieste, Italy, for the hospitality.

**Figure caption**

Figure 1(color online): $T_{s0}(p)$ (taken from ref. [29]) for Eu-LSCO (blue squares) and Y123 (green diamonds), and the oxygen isotope exponent, $\alpha(p)$ (taken from refs. [10, 15, 32]) for pure LSCO (red circles). The full red curve is drawn as a guide to the eyes.

Figure 1

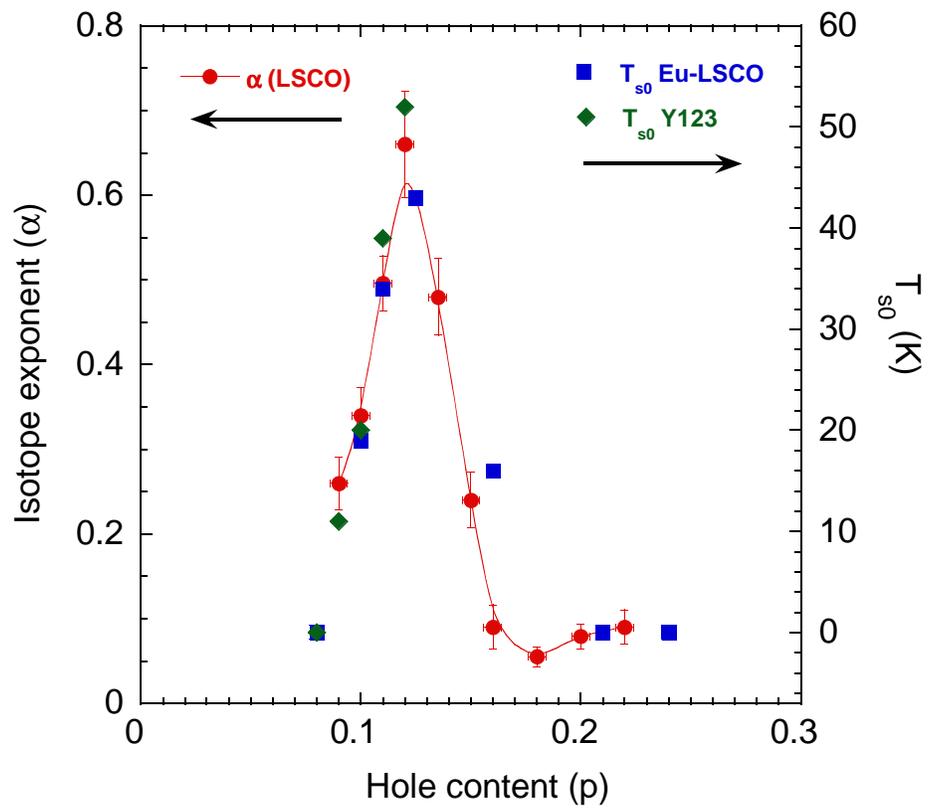